\documentclass{article}

\usepackage{amssymb,dsfont,latexsym}

\newtheorem{thm}{Theorem}

\newtheorem{prop}[thm]{Proposition}
\newtheorem{lem}[thm]{Lemma}
\newtheorem{cor}[thm]{Corollary}

\newcommand\enu[1]{\smallskip\newline\makebox[5mm][l]{\rm(#1)}}

\newcommand\bp{\noindent{\it Proof.}\ }

\newcommand{\C}{\mathbb{C}}

\begin{document}

\author{Erling St{\o}rmer}

\date{06-21-2012 }

\title{Separable states and the SPA of a positive map}

\maketitle

We introduce a necessary condition for a state to be separable and apply this condition to the SPA of an optimal positive map and give a  proof of the fact that the SPA need not be separable.

\section*{ Introduction}

It is often very difficult to verify whether a given state on a tensor product $M_n\otimes M_m$, $M_n$ being the nxn matrices, is separable.  There are necessary conditions like the PPT condition 
which are reasonably easy to verify, but no such sufficient conditions.  In the present paper we introduce two necessary conditions which are very easy to verify, and then apply them to show that the SPA of some optimal maps introduced by Ha and Kye \cite {HK1} do not correspond to separable states, yielding a counter example to the conjecture in \cite {K} stating that the SPA of an optimal unital map is separable.  It was announced in \cite {HK1} that some of their maps yield counter examples of the conjecture and that proofs will appear in \cite {HK2}.

Another consequence of our analysis is that if $\phi\colon M_n \to M_n$ is positive, then the map $\phi(1)Tr + \phi$ is super-positive, i.e. entanglement breaking, a result also shown in \cite {S2}.

Let us recall some terminology.  A linear map  $\phi\colon M_n \to M_m$ is positive, written $\phi\geq 0$, if $\phi(a)\geq 0$ whenever $a\geq 0$. $\phi$ is completely positive if it is the sum of maps of the form $AdV$ defined by $AdV(a) = V^*aV$, where V is a linear operator from $\C^m$ to $\C^n$, see \cite {C}.  $\phi$ is called optimal if there exists no completely positive map $\psi \colon M_n \to M_m$ such that $\phi - \psi \geq 0$.  If $\phi $ is unital, i.e. $\phi(1)= 1$, then the structural physical approximation, the SPA of $\phi$, is the Choi matrix (also called entanglement witness) of a completely positive map closely associated with $\phi$, see Section 2.  We recall that if $(e_{ij})$ is a complete set of matrix units for $M_n$ then the Choi matrix $C_{\phi}$ for $\phi$ is the matrix $\sum e_{ij} \otimes \phi(e_{ij}) \in M_n \otimes M_m$.  Then $\phi$ is completely positive if and only if $C_\phi$ is positive, see \cite {C}.  $\phi$ is called super-positive, or entanglement breaking, if $C_\phi$ is separable, i.e. a sum of the form $C_\phi = \sum a_i \otimes b_i$ with $a_i, b_i \geq 0$.

\section { Separable states}

Let $(e_{ij})$ be a complete set of matrix units for $M_n$.  Then the set $(e_{ij} \otimes e_{kl})_{ijkl}$ is a complete set of matrix units for $M_n\otimes M_n$.  Hence, if $a\in M_n\otimes M_n$ then $a$ is of the form

$$
a= \sum_{ijkl} a_{(ij)(kl)} e_{ij} \otimes e_{kl}.
$$
We introduce the following notation.

$$
S(a) = \sum a_{(ij)(ij)},\\
T(a) = \sum a_{(ij)(ji)}.
$$
 S(a) and T(a) give necessary conditions for a state with density operator $a$ to be separable.  Indeed we have:
 
 \begin{prop}\label{prop 2.1}
 If $a$ is a separable density operator in $M_n \otimes M_n$ then $0\leq S(a) \leq 1,  0\leq T(a)\leq 1$
 \end{prop}
 \bp
 We first consider the case when $a = b \otimes c$ with $b, c \geq 0$.  Since $Tr(a) \leq 1$ we may scale $b$ and $c$ so $Tr(b), Tr(c) \leq 1$.  Let $b= (b_{ij}), c= (c_{kl})$.  Then
 $$
 a_{(ij)(kl)} = b_{ij}c_{kl}.
 $$
 Hence
 $$
 T(a) = \sum b_{ij}c_{ji}.
 $$
  Since $bc = \sum b_{ij} c_{jk} e_{ik}$, we have
 $$
 0\leq Tr(bc) = \sum b_{ij}c_{ji} = T(a).
 $$
 Hence, if we write $Tr$ for $Tr \otimes Tr$
 $$
 T(a) \leq \parallel b \parallel Tr(c) \leq Tr(b)Tr(c) = Tr(b\otimes c) \leq 1.
 $$
 In the general case $a=\sum b^k\otimes  c^k$ with $b^k, c^k \geq 0$. Thus
 $$
 0\leq T(a) = \sum T(b^k \otimes c^k) \leq \sum Tr(b^kÊ\otimes c^k) = Tr(a) = 1.
 $$
 The proof for S(a) is similar replacing $c^k$ by its transpose $(c^k)^t$.  The proof is complete.
 
 In order to study separability Werner \cite {W} considered automorphisms of the form $AdU\otimes U$ on $M_n\otimes M_n$.  He showed the following result; for a detailed proof see Section 7.4 in \cite {S2}.
 
 \begin{lem}\label{lem 2.2}
 Let $a\in M_n\otimes M_n$ be invariant under all automorphisms $AdU\otimes U$ with $U$ unitary in $M_n$.  Then $a$ is of the form $a= \alpha 1\otimes 1 + \beta V$ with $\alpha, \beta\in \C$ and $V$ the flip $b\otimes c \to c\otimes b$ on $M_n\otimes M_n$.
 \end{lem}
 
 Let G denote the compact group $G=\{ AdU\otimes U: U \in  M_n unitary\}$. Let $dU$ be the normalized Haar measure on G.  Then
 $$ 
 P(a) = \int_G AdU\otimes U (a) dU
 $$
 is the unital trace preserving projection of $M_n\otimes M_n$ onto the fixed point algebra of G, which by Lemma 2.2 is the linear span of $1\otimes 1$ and $V$.
 
 \begin{prop}\label{Prop 2}
Let $a\in M_n\otimes M_n$.  Then
$$
T(a) = Tr(aV) = Tr(P(a)V).
$$
 \end{prop}
 \bp
 We have
 \begin{eqnarray*}
 Tr(P(a)V) &=& \int_G  Tr(AdU\otimes U (a) V)dU  \\
 &=& \int_G Tr(a AdU^*\otimes U^* (V))dU  \\
 &=& \int_G  Tr(aV)dU\\
 &=& Tr(aV).
 \end{eqnarray*}
 To show the other equality we first note that $V=\sum e_{ij}\otimes e_{ji}$.  Indeed, if $\xi_i$ is an orthonormal basis for $\C^n$ such that $e_{ij}\xi_k = \delta_{jk} \xi_i$, then
 $$
 e_{ij}\otimes e_{ji}(\xi_kÊ\otimes \xi_l) = \sum \delta_{jk}\xi_i \otimes \delta_{il}\xi_j = \xi_l\otimes \xi_k,
 $$
 proving the assertion.  If $a=\sum a_{(ij)(kl)} e_{ij}\otimes e_{kl}$,  we get
 \begin{eqnarray*}
 Tr(aV) &=& Tr(\sum_{ijkl} a_{(ij)(kl)} e_{ij} \otimes e_{kl} \sum_{rs} e_{rs} \otimes e_{sr} )\\
 &=& Tr( \sum_{j=r,l=s} a_{(ij)(kl)} e_{is} \otimes e_{kr} )\\
 &=& Tr( \sum_{ijkl} a_{(ij)(kl)} e_{il} \otimes e_{kj} )\\
 &=& \sum_{ij} a_{(ij)(ji)}\\
 &=& T(a),
 \end{eqnarray*}
proving the proposition.
 
 \begin{thm}\label{thm4}
 Let $a$ be a density matrix in $M_n\otimes M_n$, and let $\psi$ denote the positive map of $M_n$ into $M_n$ given by
 $$
 \psi(b) = \frac{n(n^2 -1)}{nT(a)-1)}  Tr(a) + a^t,
 $$
 t being the transpose map.  Then we have:
 \enu{i}  $C_{\psi}  = \frac{n(n^2 -1)}{nT(a) -1} P(a)$.
 \enu{ii}  $cTr + t $ is super-positive if and only if $c\geq 1$.
 \enu{iii} $P(a)$ is separable if and only if $T(a)\leq 1$.
 \end{thm}
 
 \bp
  By Lemma 2  $ P(a) = \alpha 1\otimes 1 + \beta V$, hence
  $$
Tr(a) = Tr(P(a)) = \alpha n^2 + \beta n.
  $$  
  By Lemma 3, since $Tr(V) = n,$
  $$
  T(a)= Tr(P(a)V)= Tr((\alpha 1\otimes1 + \beta V)V) = Tr(\alpha V + \beta 1\otimes 1) = Ê\alpha n + \beta n^2.
  $$
  Solving the last two equations for $\alpha$ and $\beta$ we get
  $$
  \alpha = \frac{n-T(a)}{n(n^2 -1)}, \,  \beta = \frac{nT(a) - 1}{n(n^2 -1)}
  $$
 Hence
 $$
 P(a) =   \frac{n-T(a)}{n(n^2 -1)}1\otimes 1 +  \frac{nT(a) - 1}{n(n^2 -1)}V.
 $$
 Since $C_{Tr} = 1\otimes 1$, and $C_t=V$ for the transpose map t, we find
 $$
 \frac{n(n^2 -1)}{nT(a) - 1} P(a) = \frac{n-T(a)}{nT(a) -1} 1\otimes 1 +V = C_{\psi},
 $$
 proving (i).
 
 In the special case when $a=e\otimes e$ with $e$ a 1-dimensional projection we have $T(a)= 1$, so by (i)
 $$
 n(n+1)P(a) = C_{\psi} = \frac{n-1}{n-1} 1\otimes 1 + V = 1\otimes 1+V,
 $$
 so $\psi = Tr + t$.  Thus, if $ \phi $ is a positive map, then
 $$
 Tr(C_{\psi} C_{\phi}) = n(n+1) \int_G Tr( AdU(e)\otimes AdU(e) C_{\phi})dU  \geq 0,
 $$
 since $Tr(b\otimes c \, C_{\phi})\geq 0$ for all $b\geq 0, c\geq 0$ when $\phi$ is a positive map.
 Thus $Tr + t $ is in the dual cone of all positive maps, so that $Tr + t$ is super-positive, see 
 \cite {S1}.  If $c\geq 1$ then 
 $$
 cTr + t = (c-1)Tr + (Tr + t)
 $$
 is the sum of two super-positive maps, so is super-positive. If $\iota$ denotes the identity map on $M_n$, then $Tr + \iota = (Tr + t)\circ t.$  Since the composition of a super-positive map and a positive map is super-positive, $Tr + \iota$ is super-positive, and as above $cTr + \iota$ is super-positive for $c\geq 0$.
 
 To complete the proof of (ii) let $0<c<1$.  We have $C_{Tr} = 1\otimes 1$ and $C_{\iota}= \sum_{ij} e_{ij}\otimes e_{ij} = ne$ with e a 1-dimensional projection.  We get
 \begin{eqnarray*}
 Tr(C_{cTr+\iota} C_{Tr - \iota}) &=& Tr((c1\otimes 1 + ne)(1\otimes 1 - ne))\\
 &=& Tr( c1\otimes 1 - cne + ne - n^2 e)\\
 &=& cn^2 -cn +n  - n^2\\
 &=& (c-1)(n^2 -n)\\
 &<&0.
 \end{eqnarray*}
  Since $Tr - \iota$ is a positive map, it follows that  $cTr +\iota$ is not in the dual cone of the positive maps, hence is not super-positive by \cite {S1}, and since $(cTr + t)\circ t = cTr + \iota, cTr + \iota$ is not super-positve either, completing the proof of (ii).
 By (i) and (ii) we have $P(a)$ is separable if and only if 
 $$
 \frac{n-T(a)}{nT(a)-1} \geq 1,
 $$
 hence if and only if $T(a)\leq 1$, proving (iii).  The proof is complete.

 \medskip

  A consequence of the above proof is the following corollary, which is also shown in \cite {S2}.
 
 \begin{cor}\label{cor5}
 Let $\phi\colon M_n\to M_n$ be a positive map. Then the map $a \to \phi(1)Tr(a) + \phi(a)$ is super-positive.
 \end {cor}
 \bp
 $\phi(1)Tr + \phi = \phi\circ(Tr + \iota)$ is super-positive since the composition of a super-positive map and a positive map is super-positive.

\section{The SPA of a map}

Let $\phi$ be a unital map of $M_n$ into itself.  Let $W=\frac{1}{n} C_{\phi}$.  Then $Tr(W)=n$ Following \cite {CP}, for $0\leq t \leq 1$ we put 
$$
\tilde W(t) = \frac{1-t}{n^2} 1\otimes 1 + tW.
$$
Then $\tilde W(t)$ has trace 1. Write $W=W^+  - W^-$  with $W^+, W^- \geq 0$ and $W_+W^- =0$, and similarly for $C_\phi$.  By \cite {CP}, equation 14, 
$$
t^* = (1+n^2 \parallel W^-\parallel)^{-1}
$$
is the maximal t such that $\tilde W(t) \geq 0$. The SPA of $\phi$, $SPA(\phi),$ is defined as
\begin{eqnarray*}
SPA(\phi)&=&  \tilde W(t^*)\\
&=& \frac{1}{n^2} (1 - \frac{1}{1+n^2 \parallel W^-\parallel}) 1\otimes 1 +  \frac{1}{1+n^2 \parallel W^-\parallel} W\\
&=& \frac{1}{n^2(1+n^2 \parallel W^-\parallel)} (n^2 \parallel W^- \parallel 1\otimes 1 + n^2 W) \\
&=& \frac{1}{1+n \parallel C_{\phi}^- \parallel}  ( \frac{1}{n} \parallel C_{\phi}^-\parallel 1\otimes 1 + \frac{1}{n} C_{\phi}) \\
&=& \frac{1}{n+n^2 \parallel C_{\phi}^- \parallel} ( (\parallel C_{\phi}^- \parallel 1\otimes 1 + C_\phi)
\end{eqnarray*}

We have just proved the first part of 

\begin{prop}\label{prop6}
If  $\phi$ is a unital positive map of $M_n$ into itself, then
$$
SPA(\phi) = \frac{1}{n+n^2 \parallel C_{\phi}^- \parallel} (\parallel C_{\phi}^- \parallel 1\otimes 1 + C_\phi).
$$
Furthermore, $SPA(\phi)$ is the density for a state, which is separable if 
$ \parallel C_{\phi}^- \parallel  = 1.$
\end {prop}
\bp
Clearly  $Tr(SPA(\phi)=1$, and by choice of $t^*$, $SPA(\phi)\geq 0$ This follows also from the fact that $\parallel C_{\phi}^- \parallel 1\otimes 1 + C_\phi \geq 0$.  The last statement follows from Corollary 5.

\begin{prop}\label{prop7}
If $\phi$ is unital and $SPA(\phi)$ is separable, then both 
$$
T(C_{\phi}), \,  S(C_{\phi}) \leq n + n(n-1)\parallel C_{\phi}^- \parallel.
$$
\end{prop}
\bp
By Propositions 1 and 6 $0 \leq T(SPA(\phi)) \leq 1$.  Since $T(1\otimes 1) = n$ it follows from Proposition 6 that 
$$
\parallel C_{\phi}^- \parallel n + T(C_\phi)  \leq n + n^2 \parallel C_{\phi}^- \parallel,
$$
so that
$$
T(C_\phi)\leq n + (n^2 -n)   \parallel C_{\phi}^- \parallel.
$$
The proof for $S(C_\phi)$ is similar.

\medskip

It follows that if either $S(C_\phi)$ or $T(C_\phi)$ is greater than $ n + n(n-1)\parallel C_{\phi}^- \parallel$, then $SPA(\phi)$ is entangled.  Thus to give an example of this one should look for $\phi$ such that $  \parallel C_{\phi}^- \parallel$ is small.  This will be done in the next section.

\section{An example}

Recall that a positive map $\phi$ is optimal if $\phi - \psi$ is never a non-zero positive map for $\psi$ completely positive, or equivalently $C_\phi - a$ is never the Choi matrix for a positive map when $a\geq 0$.  There is a conjecture \cite {K} that the SPA of an optimal positive map is separable.  Following the analysis of some generalizations of the Choi map of $M_3$ into itself by Ha and Kye \cite {HK1} we shall apply our previous results to give a counter example to the conjecture.  In \cite {HK1} it is stated that the same class of maps yield counter examples to the conjecture. Following their notation we let for $a,b,c\geq 0$, and $\pi\leq \theta \leq \pi, \,  \phi(a,b,c,\theta)$ be the map of $M_3$ into itself defined by

$$
\phi(a,b,c,\theta)=\left(\begin{array}{ccc}
ax_{11}+bx_{22}+cx_{33 } & -e^{i \theta} x_{12} & -e^{-i \theta}x_{13}  \\
-e^{-i \theta}x_{21}  & cx_{11}+ax_{22} +bx_{33} & -e^{i \theta} x_{23}\\
-e^{i \theta}x_{31} & -e^{-i \theta}x_{32}  & bx_{11}+cx_{22}+ax_{33}
\end{array}\right)
$$
where $x=(x_{ij}) \in M_3$  For simplicity write $\phi$ for $\phi(a,b,c,\theta)$.  Then the Choi matrix for $\phi$ is given by
$$
C_\phi=\left(\begin{array}{ccccccccc}
a&0&0&0&-e^{i \theta}&0&0&0&-e^{-i \theta}\\
0&c&0&0&0&0&0&0&0\\
0&0&b&0&0&0&0&0&0\\
0&0&0&b&0&0&0&0&0\\
-e^{-i\theta}&0&0&0&a&0&0&0&-e^{i\theta}\\
0&0&0&0&0&c&0&0&0\\
0&0&0&0&0&0&c&0&0\\
0&0&0&0&0&0&0&b&0\\
-e^{i\theta}&0&0&0&-e^{-i\theta}&0&0&0&a
\end{array}\right)
$$
The interesting information on $C_\phi$ is obtained from the 3x3 submatrix
$$
P(a,\theta)=\left(\begin{array}{ccc}
a&-e^{i\theta}&-e^{-i\theta}\\
-e^{-i\theta}&a&-e^{i\theta}\\
-e^{i\theta}&-e^{-i\theta}&a
\end{array}\right)
$$
Let $p_{\theta} = max \{ 2Re \, e^{i(\theta -\pi/3)}, 2Re \, e^{i\theta}, 2Re \, e^{i(\theta+\pi/3)}\}.$ Then $1\leq p_{\theta} \leq 2$, and $P(a,\theta)\geq 0$ if and only if $a\geq p_{\theta}$.  By \cite {HK1}, Theorem 2.2 we have:

\begin{lem}\label{lem7}
The map $\phi(a,b,c,\theta)$ is positive if and only if 
\enu{i} $a+b+c\geq p_{\theta}$, and
\enu{ii}$ a\leq 1 \Rightarrow bc\geq (1-a)^2$.
\end{lem}

By \cite {HK1}, Theorem 4.1 it follows that $\phi$ has the spanning property, hence is optimal by \cite {L}, under the following conditions:

\begin{lem}\label{lem8}
Suppose $\phi(a,b,c,\theta)$  is positive and $1<p_{\theta}<2$ then $\phi(a,b,c,\theta)$ is optimal
 if $0\leq a<1$, and $bc=(1-a)^2.$
\end{lem}

We can now give examples of optimal maps for which the SPA is not separable. Let
$$
P= \left(\begin{array}{ccc}
1&1&1\\
1&1&1\\
1&1&1
\end{array}\right)
$$

\begin{thm}\label{thm9}
There exist $a,b,c\geq 0$ and $0\leq \theta\leq \pi$ such that $\phi(a,b,c,\theta)$ is optimal while its SPA is entangled.
\end{thm}
\bp
Let $0<\epsilon\leq 1/4$.  Choose $\delta>0$ such that the following conditions hold:
\enu{i} If $\theta = \pi -  \delta$, then $1< p_{\theta} < 1+\epsilon$.
\enu{ii} $Re(-e^{i\theta}) > 1 - \epsilon$.
\enu{iii} $ \parallel P(a,\theta) - P \parallel <\epsilon$, where $a= p_{\theta} -Ê\epsilon$.

Let $b= \epsilon$ and
$$
c= \frac {(1-a)^2}{b} = \frac {(1-p_{\theta} + \epsilon)^2 }{\epsilon}.
$$
Since $1<p_{\theta} < 1+\epsilon,\,  0< 1-p_{\theta} + \epsilon<\epsilon.$ Thus $0<c<\epsilon$. We have
$$
a+b+c = p_{\theta} -\epsilon + \epsilon + c> p_{\theta} > 1.
$$
Thus by Lemmas 7 and 8 $\phi = \phi(a,b,c,\theta)$ is an optimal map. Furthermore
$$
1<\phi(1) = (a+b+c)1 < p_{\theta} -\epsilon +\epsilon + \epsilon = p_{\theta} + \epsilon < 1+2\epsilon,$$
and so $\psi = \parallel \phi \parallel ^{-1} \phi$ is unital.

Since $a= p_{\theta} -\epsilon < p_{\theta}$ we have, as remarked above, that $P(a,\theta)$ is not positive, and by (iii) its minimal eigenvalue lies in the interval $(-\epsilon,0)$. It follows that the minimal eigenvalue for $C_\phi, -\parallel C_{\phi}^- \parallel$, satisfies
$$
 \parallel C_{\phi}^- \parallel <Ê\epsilon.
 $$
 We now show that the conclusion of Proposition 7 is false for $\psi$ and hence that $\phi$ is not separable.  From the form of  $C_{\phi}$ we compute, denoting for simplicity of notation $k=  \parallel \phi \parallel ^{-1}$, so $\psi=k\phi$ with $(1+ 2\epsilon)^{-1}  < k < 1$.
 \begin{eqnarray*}
 S(C_{\psi})&=& kS(C_{\phi})\\
 &=& k(3a + 3 \centerdot 2 Re(-e^{i(\pi - \delta)}))\\
& >& 3k(a+2(1-\epsilon)),\,  by (ii),\\
 &=& 3k(p_{\theta} -\epsilon + 2 - 2\epsilon)\\
 &>& 3k(1+2-3\epsilon)\\
 &=& 9k(1-\epsilon)\\
 &>& 9 \frac{1-\epsilon}{1+2\epsilon}\\
 &\geq& 9/2 \\
 &>& 3 + 3 \centerdot 2  \parallel C_{\phi}^- \parallel\\
 &>&  3 + 3 \centerdot 2  \parallel C_{\psi}^- \parallel\\ ,
 \end{eqnarray*}
where we used that  $ \parallel C_{\phi}^- \parallel <\epsilon \leq 1/4$. Thus by Proposition 7 $SPA(\psi)$ is not separable, hence the same holds for $SPA(\phi)$. The proof is complete.

Department of Mathematics,  University of Oslo, 0316 Oslo, Norway.

e-mail  erlings@math.uio.no


\begin{thebibliography}{999}


\bibitem{C}
M-D. Choi,  {\em Completely positive linear maps on complex matrices}, Linear Alg. and Applic. 10(1975), 285-290.

\bibitem{CP} 
D. Chruscinski and J. Pytel, {\em Optimal entanglement witnesses from generalized reduction and Robertson maps}, J.Phys.A, 44 (2011), 165304.

\bibitem{HK1}
K-C. Ha and S-H. Kye, {\em Entanglement witnesses arising from Choi type positive linear maps}, arXiv 1205.2921(quant-ph), May 2012.

\bibitem{HK2}
 K-C. Ha and S-H. Kye, {\em The structural physical approximations and optimal entanglement witnesses}, preprint.
 
 \bibitem{K}
 J.K. Korbicz, M.L. Almeida, J. Bae, M. Lewenstein, and A. Acin, Phys.Rev.A, 78 (2008), 062105.
 
 \bibitem{L}
 M. Lewenstein, B. Kraus, J.I. Cirac, and P. Horodecki, Phys.Rev.A, 62 (2000), 052310.
 
 \bibitem{S1}
 E. St{\o}rmer, {\em Duality of cones of positive maps}, Munster J. Math. 2 (2009), 299-309.
 
 \bibitem{S2}
 E. St{\o}rmer, {\em Positive linear maps of operator algebras}, Springer-Verlag, to appear.
 
 \bibitem{W}
 R.F. Werner, {\em  Quantum states with Einstein-Podolsky-Rosen correlation admitting a hidden-variable  model}, Phys.Rev.A. 40 (1989), 4277-4281.
 

\end{thebibliography}
\end{document}